\documentclass[12pt,a4paper]{article}
\usepackage{epsfig}
\usepackage{psfig}

\begin{document}
\begin{flushright}TAUP 2668-2001\end{flushright}
\begin{center}

{\LARGE Relativistic Mechanics of Continuous Media \\} 
{\large S. Sklarz\\}
Department of Chemical Physics, Weizmann Institute of Science,\\
Rehovot 76100, Israel\\
shlomo.sklarz@weizmann.ac.il\\

{\large L. P. Horwitz\\}
Raymond and Beverly Sackler Faculty of Exact Science,\\
School of physics, Tel Aviv University,\\
Ramat Aviv 69978, Israel \footnote{Also at Department of Physics Bar Ilan University Ramat Gan 52900, Israel}
\end{center}

In this work we study the relativistic mechanics of continuous media on a fundamental level using a manifestly covariant proper time procedure. We formulate equations of motion and continuity (and constitutive equations) that are the starting point for any calculations regarding continuous media. In the force free limit, the standard relativistic equations are regained, so that these equations can be regarded as a generalization of the standard procedure.
In the case of an inviscid fluid we derive an analogue of the Bernoulli equation. For irrotational flow we prove that the velocity field can be derived from a potential. If in addition, the fluid is incompressible, the potential must obey the d'Alembert equation, and thus the problem is reduced to solving the d'Alembert equation with specific boundary conditions (in both space and time). The solutions indicate the existence of light velocity sound waves in an incompressible fluid (a result known in previous literature \cite{Synge276}).
Relaxing the constraints and allowing the fluid to become linearly compressible one can derive a wave equation, from which the sound velocity can again be computed. For a stationary background flow, it has been demonstrated that the sound velocity attains its correct values for the incompressible and nonrelativistic limits. Finally viscosity is introduced, bulk and shear viscosity constants are defined, and we formulate equations for the motion of a viscous fluid.

%%%%%%%%%%%%%%%%%%%%%%%%%%%%%%%%%%%%%%%%%%%%%%%%%%%%%%%%%
\section{Introduction}

We wish to formulate manifestly covariant equations for the motion of a continuum on a fundamental level, and to demonstrate their use in an interesting problem. In attempting to reach a consistent theory one meets a few difficulties:

The general motion of a continuum may include accelerations, making the use of standard special relativistic dynamics  problematic. The usual procedure for dealing with an accelerating particle, would be to go from one inertial frame to the next, as the particle accelerates, always keeping it instantaneously in an inertial frame \cite{Weinberg}. In the case of a continuum however, this procedure is impossible, for each of  the infinitesimal volume elements accelerates, in general, at different rates and directions, and so only one specific portion of the volume can be kept at rest at any particular time. Secondly the above procedure may not be manifestly covariant, as an accelerating particle momentarily at rest in some frame, is obviously not equivalent to an identical particle moving at constant speed. The accelerating particle, for instance, will be radiating if it is charged. Our claim is that keeping a particle momentarily at rest can, at best, yield an approximation to the real motion, but is by no means a consistent covariant treatment of the system.

Another severe difficulty in writing down relativistic equations of hydrodynamics lies in the relativistic property of simultaneity. 
In order to define an extended body in mathematical terms, a certain configuration of the body is chosen arbitrarily and is referred to as the reference configuration. In the nonrelativistic theory, each particle, or infinitesimal volume element of the body, can then be labeled throughout the evolution by its position in that reference configuration.  The difficulty in constructing a relativistic theory of hydrodynamics lies in the fact that each inertial system will have its own notion of time and simultaneity, and they  will not be able to agree on any set of particles to be considered as existing simultaneously at any given time $t$. It therefore becomes impossible to define a consistent reference configuration.

In 1941 Stueckelberg \cite{Stueckelberg} postulated that {\it events} (not particles) are to be considered as the fundamental dynamical objects of the theory of motion. The dynamics of an event is governed by 8 independent parameters consisting of the space-time coordinates\footnote{Here and throughout this work we use the usual convention where Lattin indices $i, j, k, ...$ take the values 1, 2, 3, related to Euclidean space, whereas Greek indices $\mu, \nu, ...$ take the values 0, 1, 2, 3 and correspond to the four dimensions of Minkowski space-time} $x^\mu=(ct,{\bf r})$ and energy momenta $p^\mu=(E/c,{\bf p})$ associated with each event, and parameterized by an invariant time. Based on this, Horwitz and Piron in 1973 proposed that {\it there exists a universal time $\tau$, by means of which dynamical interactions are correlated} \cite{Horwitz92}. This universal time may be identified with the Robertson-Walker time of the expansion of the universe, the time read on a freely falling ideal clock.  The evolution of a system of $N$ events through $\tau$ is generated by an invariant function $K$ on this $8N$ dimensional phase space, by means of the covariant Hamilton equations \cite{SyngeGas}:

\begin{equation} \label{eintro1}
   \frac{dx_i^{\mu}}{d\tau} = \frac{\partial K}{\partial p_{i\mu}}, \qquad
   \frac{dp_i^{\mu}}{d\tau} = -\frac{\partial K}{\partial x_{i\mu}}, 
\end{equation}
The collection of events along each world line corresponds to a particle, and hence the evolution of the state of the $N$-event system describes,{\it  a posteriori}, the history in space and time of an $N$-particle system. 
In the case of a free particle, for instance, we choose for the Hamiltonian:

\begin{equation} \label{eintro2}
   K = \frac{p_\mu p^\mu}{2M},
\end{equation} 
where $M$ is a positive parameter, which is a given intrinsic property of an event, having the dimension of mass and emerging as the Galilean target mass in the nonrelativistic limit $c\to \infty$.  This results in 

\begin{eqnarray} \label{eintro3} 
   \frac{dx^\mu}{d\tau}=\frac{p^\mu}{M}, \qquad
   \frac{dp^\mu}{d\tau}=0,
\end{eqnarray}
so that we have for the physical velocity (we take $c=1$ in what
follows unless otherwise specified),

\begin{equation} \label{eintro4}
   {\bf v}=\frac{d{\bf x}}{dt}=\frac{{\bf p}}{E}
\end{equation} 
in total agreement with the usual Einstein kinematics.
As $(E,{\bf p})$ are each taken to be independent variables of the motion, the particles are not necessarily on mass shell, and the particle mass squared: $m^2=E^2-p^2$, is not contrained to be constant (except for a free particle), but is rather to be determined as a solution of the dynamical equations. Note that (we use the metric $-+++$)

\begin{equation} \label{eintro5}
   -\frac{dx_\mu}{d\tau}\frac{dx^\mu}{d\tau}=\frac{ds^2}{d\tau^2}
                                            =\frac{m^2}{M^2}
\end{equation} 
and is unity only for the ``mass-shell'' value $m^2=M^2$, Hence $\tau$ corresponds to the proper time of an ideal  free clock on its mass shell.
On mass shell, the time obeys the relation:

\begin{equation} \label{eintro6}
   \frac{dt}{d\tau}=\frac{E}{M}=\frac{1}{\sqrt{1-v^2}},
\end{equation}
so that $dt$ is precisely the time interval measured in the laboratory between two signals emitted by a source, traveling with velocity $v$, with interval $d\tau$, according to the Lorenz transformation. If the emitter is not on shell there is a factor $m/M$. The red shift imposed by general relativity, $\Delta t=\sqrt{-1\over g_{00}}\Delta\tau$ is another example of the effect of dynamics, in this case, the acceleration due to gravity (contained in $g_{00}$), on the observed time interval. The observed time interval is therefore influenced by forces \cite{Horwitz88} that may move the energy momentum off shell. 

This theory has since been applied to fundamental questions such as the Newton-Wigner position operator and the Landau-Peierls relation \cite{Arshansky85}, and to dynamical problems such as two body bound states \cite{Arshansky88,Arshansky89} and scattering and electromagnetic interactions in classical mechanics  \cite{Oron,Land91}, quantum mechanics, and quantum field theory \cite{Shnerb93,Land95}.

In order to formulate equations of motion for continuous media,
we therefore turn to the above mentioned procedure of  Stueckelberg, Horwitz and Piron \cite{Stueckelberg}, and all following discussions will be conducted in the framework of this theory. As a consequence of the existence of a universal time which parameterizes all systems, it becomes possible to define a certain configuration of events in space-time $X^\mu$, at some definite initial universal time $\tau_0=0$, and then follow the evolution of the system through $\tau$. The universality of $\tau$ eliminates any ambiguity as to the identification of events and their distribution in space-time. Labeling the various events by their initial configuration $X^\mu$, one can thus express their respective positions at any later time $\tau$:
 
\begin{equation}\label{eintro7}
  x^\mu=x^\mu(X^\nu,\tau).
\end{equation} 
The velocity and acceleration of an event labeled $X^\mu$  at time $\tau$ is thus:

\begin{eqnarray}
V^\mu(X^\nu,\tau)&=&\frac{dx^\mu(X^\nu,\tau)}{d\tau}\\
A^\mu(X^\nu,\tau)&=&\frac{dV^\mu(X^\nu,\tau)}{d\tau}=\frac{d^2x^\mu(X^\nu,\tau)}{d\tau^2}
\end{eqnarray} 
The Eulerian velocity field describes the velocities of events passing a certain point in space-time, as a function of the universal time:

\begin{equation}
    v^\mu(x^\gamma,\tau)=V^\mu(X^\nu(x^\gamma,\tau),\tau)
\end{equation}
We now wish to express the acceleration in terms of the Eulerian velocity field:

\begin{eqnarray}
A^\mu(X^\nu,\tau)&=&\frac{dV^\mu(X^\nu,\tau)}{d\tau}=\frac{dv^\mu(x^\gamma(X^\nu,\tau),\tau)}{d\tau}\\
a^\mu(x^\gamma,\tau)&=&A^\mu(X^\nu(x^\gamma,\tau),\tau)\nonumber \\
&=&\frac{\partial v^\mu}{\partial \tau}+\frac{\partial x^\nu}{\partial \tau}\frac{\partial v^\mu}{\partial x^\nu}
=\frac{\partial v^\mu}{\partial \tau}+v^\nu\frac{\partial v^\mu}{\partial x^\nu}.
\end{eqnarray}

In space-time, due to mass-energy equivalence, there is no {\it a priori} conservation of mass of individual particles; rather we shall be dealing with a ``conservation of events''. Let $n(x^\mu,\tau)$, be the density of events (this quantity is a scalar since the four volume element is invariant) at the space-time point $x^\mu$ at time $\tau$ multiplied by $M$, the intrinsic, given mass dimension property attributed to each event. Consider now a small volume of fluid. The flux of events through a closed surface enclosing the volume must cause a decrease in the density of events within the volume. It is in place to stress at this point that the surface mentioned is a three dimensional surface, the direction of which is defined by a four vector normal to that surface in four dimensional space-time. Hence, for example, we define $dS^0$ as a volume of three dimensional space, its direction being the time direction. Likewise, the volume over which the integration is carried out is obviously the four dimensional space-time. The flux of events past an infinitesimal surface $dS^\mu$ is: $\varepsilon(v)nv^\mu dS_\mu$, where $\varepsilon (v)=\pm 1$ depending as to whether $v^\mu$ is space-like ($+1$) or time-like ($-1$). The reason for introducing this additional sign is that for $v^\mu$ and $dS^\mu$ both pointing out of the bounded volume, in a time-like direction, the vector product $v^\mu dS_\mu$ is negative whereas the flux of events must be positive \cite{Synge276}. Therefore:

\begin{equation}
-\frac{\partial}{\partial \tau}\int_V ndV=\int_S\varepsilon(v)nv^\mu dS_\mu
\end{equation}
or, using the four-dimensional Gauss law, \cite{Synge}

\begin{equation}
\int_V\frac{\partial nv^\mu}{\partial x^\mu}dV=\int_S\varepsilon(v) nv_\mu dS^\mu,
\end{equation}
we have,

\begin{equation}
-\int_V\frac{\partial n}{\partial \tau}dV=\int_V\frac{\partial nv^\mu}{\partial x^\mu}dV
\end{equation}
and so taking the volume to be infinitely small:

\begin{equation}\label{eintro20}
\frac{\partial n}{\partial \tau}+\frac{\partial nv^\mu}{\partial x^\mu}=0,
\end{equation}
which is the equation of continuity for the flow field.
Note that in the case where the density of events $n$ is constant, the above equation reduces to:
\begin{equation}\label{eintro21}
\partial_\mu v^\mu=0.
\end{equation}

In a continuum we focus on an infinitesimal volume element and describe the forces acting on it. In general one talks about two kinds of forces:
\begin{enumerate}

\item Forces which are proportional to the density of mass, charge or number of events, within the volume. The force will therefore be $nb^\mu$, where $n$ is the density of events.

\item  Surface contact forces, which are exerted by the continuum of  events surrounding the volume under consideration, and are transferred through its surface. The number of events neighboring a small portion of the surface, $dS$, in a locally homogenous fluid, are obviously proportional to its size, hence we can assume the force to be proportional to the surface it is acting upon. In mathematical terms we assert that the force $df$  on an element of surface $dS$ tends to zero as $dS$ tends to zero, but that the ratio $df/dS$ tends to a definite limit. This assumption enables us to define a stress four tensor, the components of which describe the force in the $\mu$ direction, exerted on a surface in the $\nu$ direction: 

\begin{equation}
\sigma^{\mu\nu}=\frac{df^\mu}{dS_\nu}.
\end{equation}
The surface force on an elementary volume unit, in the $\mu$ direction is therefore (by the usual argument): 	
\begin{equation}
nf^\mu=\frac{\partial\sigma^{\mu\nu}}{\partial x^\nu}
\end{equation}
\end{enumerate}

Summing up the forces acting on an infinitesimal volume element and equating to the density of acceleration within the volume, we get the equations of motion:
\begin{equation}\label{eintro24}
\frac{\partial\sigma^{\mu\nu}}{\partial x^\nu}+nb^\mu=n\left(\frac{\partial v^\mu}{\partial \tau}+v^\nu\frac{\partial v^\mu}{\partial x^\nu}\right)
\end{equation}

By introducing a manifestly covariant Boltzmann equation, Horwitz, Shashoua and Schieve \cite{boltzmann} were able to write down equations for conserved quantities in a statistical mechanical framework. It is interesting to note that the equations of continuity and motion derived above from a hydrodynamical point of view are equivalent to those presented by these authors for the conservation of some constant quantity and for the conservation of momentum respectively.  

In order to make these equations of motion operational it is necessary to formulate constitutive equations for the fluid. These will define the way by which the stress forces acting at a certain point in the fluid $\sigma^{\mu\nu}$, are governed by its physical state (and/or history). These laws however are determined by physical properties of the material; we shall make some simplifying assumptions about the nature of the fluid, comparable to those which lead to the Navier-Stokes equations in the non-relativistic case. 

In the following work we shall formulate a potential theory for inviscid, irrotational, incompressible fluids. Then, relaxing the constraints, we shall allow the fluid to become linearly compressible, an assumption which will lead to acoustic modes within the flow. Finally we shall introduce equations for the motion of viscous flow.

%%%%%%%%%%%%%%%%%%%%%%%%%%%%%%%%%%%%%%%%%%%%%%%%%%%%%%%%%
\section{Standard relativistic limit}

In the current section we shall show how Eq. (\ref{eintro20}) and (\ref{eintro24}) transform into the standard relativistic equations of hydrodynamics, which in turn give the nonrelativistic limit by taking $c\to\infty$. We begin with a few preliminary remarks which will assist us in what follows. Note the following connection between the relativistic force term $f^\mu$ and its nonrelativistic counterpart ${\bf f}^i$:

\begin{eqnarray}
f^\mu &=& M\frac{d^2 x^\mu}{d\tau^2}=M\frac{d}{d\tau}\left(\frac{dx^\mu}{dx^0}\frac{dx^0}{d\tau}\right)\nonumber\\
&=& M \left[\frac{dx^0}{d\tau}\frac{d}{d\tau}\left(\frac{dx^\mu}{dx^0}\right)+
           \frac{dx^\mu}{dx^0}\frac{d^2 x^0}{d\tau^2}\right]\nonumber\\
&=& M \left[\left(\frac{dx^0}{d\tau}\right)^2\frac{d^2x^\mu}{dx^{02}}+
           \frac{dx^\mu}{dx^0}\frac{d^2 x^0}{d\tau^2}\right]
\end{eqnarray}
Taking the space components we can write

\begin{equation}\label{elimit0}
f^i=\left(\frac{v^0}{c}\right)^2{\bf f}^i+{\bf v}^if^0,
\end{equation}
A rearrangement of terms gives the physically measureable force ${\bf f}^i$ in terms of the relativistic force,

\begin{equation}\label{elimit1}
{\bf f}^i=\left(\frac{c}{v^0}\right)^2(f^i-{\bf v}^i f^0).
\end{equation}
Another useful expression which connects derivatives of the velocity field $v^\mu$ with its nonrelativistic physically observable counterpart ${\bf v}^i=cv^i/v^0$, will be used later and can be proved as follows (we denote by $\partial_s$ a derivative with respect to any variable):

\begin{equation}\label{elimit2}
\partial_s v^i=\partial_s (\frac{v^0 {\bf v}^i}{c})
=\frac{v^0}{c}\partial_s{\bf v}^i+\frac{{\bf v}^i}{c}\partial_s v^0.
\end{equation}

We remind ourselves too, that $v^0=dx^0/d\tau=E/M$, and that in the nonrelativistic (on-shell) limit $v^0=c/\sqrt{1-\frac{v^2}{c^2}}\approx c(1+\frac{1}{2}\frac{v^2}{c^2})$. 

To put our equations into correspondence with standard results, we must note that any interaction with an apparatus or some macroscopical object corresponds to an interaction with a worldline, an integration over $\tau$ is therefore necessary. In performing this procedure to obtain the on shell limit, we assume that correlations in $\tau$ survive only over a short interval, thus a term of the form: 
${1\over \Delta T} \int a(\tau)b(\tau)d\tau$ can be factored to obtain $\approx{1 \over \Delta T^2}\int a(\tau)d\tau \int b(\tau')d\tau'=ab$ in a $\tau$ averaged sense. \cite{Shnerb93}.
\footnote{ If both $a(\tau)$ and $b(\tau)$ have support only arround the zero mode then 
\begin{eqnarray}
   {1\over \Delta T} \int a(\tau)b(\tau)d\tau &=& {1\over \Delta T}\int \tilde{a}(s)\tilde{b}(-s)ds \nonumber\\ 
   &\approx&{1\over \Delta T^2} \tilde{a}(0)\tilde{b}(0)\nonumber\\
   &\approx&{1 \over \Delta T^2}\int a(\tau)d\tau \int b(\tau')d\tau'
\end{eqnarray}
}

Applying this procedure to equations (\ref{eintro20}) and (\ref{eintro24}) causes the $\tau$ derivative terms in both equations to drop due to the assumption that both $n$ and $v^\alpha$ vanish pointwise as $\tau\to\infty$.  
From Eq. (\ref{eintro20}) we then get:

\begin{eqnarray}
0 &=& \partial_0(nv^0)+ \partial_i(nv^i) \nonumber \\
&=& \frac{\partial}{\partial t}(nv^0)+ \nabla\cdot({\bf v} nv^0) \label{elimit3a}\\
&=&v^0\left(\frac{\partial n}{\partial t}+\nabla\cdot(n{\bf v})\right)+n\left(\frac{\partial}{\partial t}+{\bf v}\cdot\nabla\right)v^0 \label{elimit3}
\end{eqnarray}
where all quantities are from now to be understood as being $\tau$ averaged.

Written in the form (\ref{elimit3a}), Eq. (\ref{eintro20}) can be interpreted as expressing the conservation of energy density (or the continuity of energy flow), whereas the second form (\ref{elimit3}) states that the conservation of mass is to be corrected by a relativistic term. This can be seen by noting that the first bracketed term in (\ref{elimit3}), is just the regular nonrelativistic equation of continuity which gives the change in time of the mass of a small volume of three dimensional space. In nonrelativistic hydrodynamics this term should obviously be zero, for the mass is conserved. This however is not the case for a relativistic fluid, as an increase in the energy also has to be taken into account in the conservation law. This is precisely the  meaning of the second term. It expresses the change of energy of the small volume of fluid moving with the particle. This change of energy causes a corresponding non vanishing change of mass which must balance it out in such a way that the total be zero. It should be noticed that the derivatives of $v^0$ which appear in the second term of (\ref{elimit3}) are in the standard relativistic limit, of the order $1/c$, and so are small compared to the first term. Therefore when taking the nonrelativistic limit this second term can be neglected, and one is left with the standard equation for the conservation of mass:
\begin{equation}\label{elimit3b}
   \frac{\partial n}{\partial t}+\nabla\cdot(n{\bf v})=0.
\end{equation}	

We shall now treat the space and time components of Eq. (\ref{eintro24}) separately. We consider a fluid free of any external forces and therefore take $b=0$. For the sake of simplicity of notation we assume here that $\sigma^{\mu\nu}=-pg^{\mu\nu}$. A physical justification for this simplification will be forthcoming in the next section, but the generality of our following discussion is not based on or restricted by it. Taking the time component we have:

\begin{eqnarray}
-\partial^0 p &=& nv^\mu\partial_\mu v^0 \nonumber\\
&=& \frac{1}{c}nv^0\left({\bf v}\cdot\nabla+\frac{\partial}{\partial t}\right)v^0, \nonumber
\end{eqnarray}
or,
\begin{equation} \label{elimit4}
\frac{\partial p}{\partial t} = nv^0\left({\bf v}\cdot\nabla+\frac{\partial}{\partial t}\right)v^0 
\end{equation}
This  gives the time change of energy moving with the fluid. Inserting the standard relativistic limit for $v^0$, and neglecting terms of order $1/c^2$ we are left with 

\begin{eqnarray} \label{elimit8}
\frac{\partial}{\partial t}p &=& n\left({\bf v}\cdot\nabla+\frac{\partial}{\partial t}\right)\frac{1}{2}v^2 \nonumber\\
&=& \left({\bf v}\cdot\nabla+\frac{\partial}{\partial t}\right)(\frac{1}{2}nv^2)-\frac{1}{2}v^2\left({\bf v}\cdot\nabla+\frac{\partial}{\partial t}\right)n \nonumber\\
&=& \frac{\partial}{\partial t}(\frac{1}{2}nv^2)+\nabla\cdot({\bf v}\frac{1}{2}nv^2),
\end{eqnarray}
where we made use in the third line of the equation of continuity (\ref{elimit3b}). Eq (\ref{elimit8}) is evidently an equation for the change in nonrelativistic kinetic energy. The first term on the right hand side can be interpreted as the change of kinetic energy within a small volume of space whereas the second term is the flux of energy leaving that volume.

Turning to the space components, and using relation (\ref{elimit2}) we get:

\begin{eqnarray} \label{elimit5}
-\partial^i p &=& nv^\mu\partial_\mu v^i \nonumber\\
&=&\frac{n}{c}(v^0 v^\mu\partial_\mu {\bf v}^i+{\bf v}^i v^\mu\partial_\mu v^0).
\end{eqnarray}
Inserting Eq. (\ref{elimit4}) and rearranging terms yields
\begin{equation}\label{elimit6}
-\left(\frac{c}{v^0}\right)^2\left(\nabla p+\frac{{\bf v}}{c^2}\frac{\partial p}{\partial t}\right) =n\left(\frac{\partial}{\partial t}+{\bf v}\cdot\nabla\right){\bf v}.
\end{equation}
By remembering relation (\ref{elimit0}) one can notice at once that the left hand side of Eq. (\ref{elimit6}) is simply the physical force acting on an infinitesimal volume of fluid, whereas the right hand side gives the rate of change of the physical velocity moving with the fluid. In fact it can clearly be seen that in the nonrelativistic limit $c\to \infty,\quad v^0/c\to 1,$  Eq. (\ref{elimit6}) simply  becomes the non-relativistic Euler's equation of hydrodynamics.
It is worth noting too, that if we accept a correspondance between the density of events $n$ and the sum of the internal energy and pressure $w=\varepsilon+p$ (also called the heat function), then Eq. (\ref{elimit6}) is precisely equivalent to the relativistic hydrodynamics equation derived by Weinberg \cite{Weinberg49} and Landau \cite{Landau126} in the forceless events framework of standard relativity for which $(v^0/c)^2=\gamma^2$.

%%%%%%%%%%%%%%%%%%%%%%%%%%%%%%%%%%%%%%%%%%%%%%%%%%%%%%%%%
\section{Potential Theory for Relativistic Hydrodynamics}

Under certain circumstances some problems in fluid dynamics can be solved in an approximation in which viscous forces are neglected, and the fluid is assumed to be incapable of sustaining shear forces. If the fluid is assumed also to be spatially and temporally isotropic, we can write down the stress tensor as follows:
\begin{equation} \label{epot1}
\sigma^{\mu\nu}=-pg^{\mu\nu}=\left\{
	\begin{array}{rrr}
                p    &{\rm when}    & \mu=\nu=0   \\
		-p   &{\rm when}    & \mu=\nu\ne 0\\
		0    &{\rm when}    & \mu\ne\nu
        \end{array}\right.
	 =\left(
	\begin{array}{cccc}
		p  & 0  & 0  &  0\\
		0  & -p & 0  &  0\\
		0  & 0  & -p &  0\\
		0  & 0  & 0  & -p
	\end{array}\right),
\end{equation}
where $g^{\mu\nu}$ is the metric tensor for flat Minkowski space and $p$ is a Lorentz scalar. The interpretation of the stress tensor components in this form, is as follows: The three space-space components relate to the force exerted within the fluid per unit surface per unit time, whereas the time-time component describes the forces acting in the time direction per unit volume of the fluid (essentially mass-changing forces).

Substituting Eq. (\ref{epot1}) into the equations of motion,

\begin{equation} \label{epot2}
\partial_\mu\sigma^{\mu\nu}+nb^\nu=n(\frac{\partial}{\partial\tau} +v^\mu\partial_\mu)v^\nu,
\end{equation}
this leads to

\begin{equation} \label{epot3}
-\partial^\nu p+nb^\nu=n(\frac{\partial}{\partial\tau} +v^\mu\partial_\mu)v^\nu.
\end{equation}
Eq. (\ref{epot3}) may be written in different form by introducing two new variables. We define $\Psi=\int^n {1\over n'}{\partial p(n')\over \partial n'}dn'$, so that ${1\over n}\partial^\mu p =\partial^\mu \Psi$. Note that for incompressible fluid $n$ is constant and we have simply $\Psi=p/n$. Assuming too that the body force can be derived from a potential function $b^\nu=-\partial^\nu\Omega$ , we can write: 

\begin{equation} \label{epot4}
-\partial^\nu(\Psi+\Omega)=(\frac{\partial}{\partial\tau} +v^\mu\partial_\mu)v^\nu,
\end{equation}

An important kinematic result which shall be used later, is known as the Circulation Theorem.
Consider a closed circuit $C$ linking a continuous line of fluid events. In general the space-time configuration of the loop $C$ depends on $\tau$. We denote this symbolically by $C_\tau$. The kinematic Theorem to be proven is :

\begin{equation} \label{epot5}
\frac{\partial}{\partial\tau}\oint_{C_\tau}\!v_\mu dx^\mu=\oint_{C_\tau}\!a_\mu dx^\mu,
\end{equation}
where  $a^\mu$ is the space-time acceleration field of the fluid.
The proof can be obtained by transformation to the material coordinates $X^\mu$ as follows:
\begin{equation} \label{epot6}
\oint_{C_\tau}\!v_\mu dx^\mu=\oint_{C_0}\!V_\nu\frac{dx^\nu}{dX^\mu}dX^\mu.
\end{equation}
So then:

\begin{equation} \label{epot7}
\frac{\partial}{\partial\tau}\oint_{C_\tau}\!v_\mu dx^\mu=
\oint_{C_0}\!V_\nu\frac{\partial}{\partial\tau}\left(\frac{dx^\nu}{dX^\mu}\right)dX^\mu
+\oint_{C_0}\!\frac{\partial V_\nu}{\partial\tau}\frac{dx^\nu}{dX^\mu}dX^\mu.
\end{equation} 
The first integral on the right hand side of Eq. (\ref{epot7}), can be shown to vanish:

\begin{eqnarray} \label{epot8}
\oint_{C_0}\!V_\nu\frac{\partial}{\partial\tau}\left(\frac{dx^\nu}{dX^\mu}\right)dX^\mu 
&=&\oint_{C_0}\!V_\nu\left(\frac{dV^\nu}{dX^\mu}\right)dX^\mu \nonumber\\
&=&\frac{1}{2}\oint_{C_0}\!\frac{d(V_\nu V^\nu)}{dX^\mu}dX^\mu
=\frac{1}{2}\left[ V_\nu V^\nu\right]^{X_1}_{X_2}=0, 
\end{eqnarray}
since the path is closed.
The second can be transformed back to the spatial coordinates:

\begin{equation} \label{epot9}
\oint_{C_0}\!\frac{\partial V_\nu}{\partial\tau}\frac{dx^\nu}{dX^\mu}dX^\mu=\oint_{C_0}\!A_\nu\frac{dx^\nu}{dX^\mu}dX^\mu=\oint_{C_\tau}\!a_\mu dx^\mu,
\end{equation}
proving the theorem.

Substituting for $a^\mu$ in Eq. (\ref{epot5}), from the equation of motion (\ref{epot4}) gives:

\begin{eqnarray} \label{epot10}
\frac{\partial}{\partial\tau}\oint_{C_\tau}\!v_\mu dx^\mu &=& -\oint_{C_\tau}\!\partial_\mu(\Psi+\Omega)dx^\mu \nonumber\\
&=& -\left[\Psi+\Omega\right]^{X_1}_{X_2}=0.
\end{eqnarray}
where it has been assumed in the last step that the functions are single valued.
This leads generally to: 
\begin{equation}\label{epot11}
\oint_{C_\tau}\!v_\mu dx^\mu={\rm Const}
\end{equation}
If the motion of the fluid was generated from rest, so that for some initial time the circulation was zero, we get:

\begin{equation}\label{epot12}
\oint_{C_\tau}\!v_\mu dx^\mu=0,
\end{equation}
for all times.
A necessary and sufficient condition, for Eq. (\ref{epot12}), with continuous $v^\mu$, is that $v^\mu$ be expressible as a gradient of a potential function. We write
\begin{equation}\label{epot13}
v^\mu=-\partial^\mu\Phi.
\end{equation}
where $\Phi$ is defined as the velocity potential.
It has been shown above (Eq. (\ref{eintro21}) that for an incompressible fluid
\begin{equation}\label{epot14}
 \partial_\mu v^\mu=0 ,
\end{equation} 
so combining the two, we find that the velocity potential, $\Phi$, satisfies the d'Alembert equation:
\begin{equation}\label{epot15}
\partial_\mu\partial^\mu\Phi=0 ,
\end{equation}
providing an enormous simplification in the theory of  inviscid irrotational flow.
Using the four dimensional Stokes theorem \cite{Eddington} we can convert the line integral in Eq. (\ref{epot12}) into a spatial integral:

\begin{equation}\label{epot16}
\oint_{C_\tau}\!v_\mu dx^\mu=\int\!\!\int_S(\partial_\mu v_\nu-\partial_\nu v_\mu)dS^{\mu\nu}=0.
\end{equation} 
If this is to be true for arbitrary surface $S$, the integrand must vanish, giving

\begin{equation}\label{epot17}
(\partial_\mu v_\nu-\partial_\nu v_\mu)=0.
\end{equation} 
Using this we can evaluate the following expression:

\begin{equation}\label{epot18}
v^\mu\partial_\mu v^\nu= v^\mu\partial^\nu v_\mu=\frac{1}{2}\partial^\nu(v_\mu v^\mu).
\end{equation}
Inserting Eq. (\ref{epot18}) into the equations of motion (\ref{epot4}), and rearranging terms, we obtain:

\begin{equation}\label{epot19}
-\partial^\nu\left(\Psi+\Omega+\frac{1}{2}(v_\mu v^\mu)\right)=\frac{\partial}{\partial\tau}v^\nu.
\end{equation}
Inserting Eq (\ref{epot13}), into (\ref{epot19}) and rearranging terms, leads to further simplification:

\begin{equation}\label{epot20}
-\partial^\nu\left(\Psi+\Omega+\frac{1}{2}(v_\mu v^\mu)-\frac{\partial}{\partial\tau}\Phi\right)=0,
\end{equation}  
which means that:

\begin{equation}\label{epot21}
\Psi+\Omega+\frac{1}{2}(v_\mu v^\mu)-\frac{\partial}{\partial\tau}\Phi=f(\tau),
\end{equation}  
where $f(\tau)$ is some arbitrary function of $\tau$. 
This can be considered a relativistic equivalent to the Bernoulli equations\footnote{By this we mean solely that this equation fulfills, the same function in the logical procedure as does the Bernoulli equation in the nonrelativistic counterpart.}.

We return now to Eq. (\ref{epot15}) and assert that under the assumptions made, our problem reduces to solution of the d'Alembert equation, given specific boundary conditions. These require that the component of the velocity field normal to the boundary vanish everywhere on its surface.  Thus the motion of the fluid will be governed solely by the space-time geometry of the problem.  The solution can then be inserted back into the equations of motion (\ref{epot21}), determining the distribution of forces and stresses within the fluid.

The general solution to the d'Alembert equation (\ref{epot15}), can be written as an integral over all possible plane waves:

\begin{equation}\label{epot22}
\Phi=\int d^4kA(k^\mu)\exp(ik_\mu x^\mu),
\end{equation}	
where $k^\mu$ is restricted to the shell 
\begin{equation}\label{epot23}
k_\mu k^\mu=0,  
\end{equation}
and the $A(k^\mu)$ which determine the shape of the wave packet are to be determined by boundary and initial conditions.

It must be noted though, that there also always exists another trivial solution, because  any solution to the Laplace equation with an additional linear time term, will also satisfy the d'Alembert equation. The space part, $\Phi_\nabla$, by definition, vanishes under the Laplacian, while the time component being linear, vanishes when differentiated twice. This trivial solution
\begin{equation}\label{epot24}
\Phi_{NR}=\Phi_\nabla+u_0x^0,
\end{equation}
gives rise to a physical velocity field:

\begin{equation}\label{epot25}
{\bf v}^i= \frac{v_{NR}^i}{v_{NR}^0}= \frac{1}{u^0}\partial^i\Phi_\nabla
\end{equation}
which is clearly to be interpreted as the nonrelativistic flow of an incompressible fluid. The total solution then should be written as a linear combination
\begin{equation}
\Phi_{TOT}=\Phi_{NR}+\Phi.
\end{equation}
As for the physical motion of the fluid we now have:

\begin{equation}
{\bf v}^i =\frac{v^i_{NR}+v^i}{v^0_{NR}+v^0}
\end{equation}
If we assume, as is reasonable for low energies, that the nonrelativistic part of the flow is far greater than the relativistic correction, $v^\mu_{NR}\gg v^\mu$, then an expansion in a Taylor series gives:

\begin{eqnarray}
{\bf v}^i &\approx& \frac{v^i_{NR}+v^i}{u^0}\left(1-\frac{v^0}{u^0}\right)\nonumber\\
&\approx& \frac{v^i_{NR}}{u^0}+ \frac{v^i}{u^0}-\frac{v^i_{NR}v^0}{(u^0)^2}.
\end{eqnarray}
The first term is just the nonrelativistic, $t$ independent flow of Eq. (\ref{epot25}), whereas the other two are of a smaller order and describe time dependent wave-like fluctuations. It seems, then, that the covariant form for the potential flow gives rise to a physical velocity which can be divided  into two parts ${\bf v}\approx {\bf v}_{NR}+{\bf v}_R$, an underlying nonrelativistic velocity field and  an additional relativistic correction with a wave-like nature, due to disturbances in the background flow. These disturbances can be envisioned as ripples riding above the usual nonrelativistic flow, at a speed which will be shown to be the speed of light. 

The velocity of the disturbances can be calculated in various ways. Writing $k^\mu=kn^\mu$ in Eq. (\ref{epot22}), and picking out a specific direction $n^\mu$, we can perform the integration over the magnitude $k$, getting:

\begin{equation}
\Phi=f(n_\mu x^\mu).
\end{equation}
But from relation (\ref{epot23}), $n_\mu n^\mu=0$, so the phase velocity $v_f$, of the wave front is:

\begin{equation}
v_f=\left|\frac{n^0}{n_i}\right|=1
\end{equation}

If the solution given above (\ref{epot22}), describes a wave packet, the group velocity of this packet can be determined through the stationary phase method. The highest contribution to the integral will come from those $k$'s for which the phase is stationary:

\begin{eqnarray}
\frac{d}{dk^i}\left(k_\mu x^\mu\right)=0 \nonumber\\
x^i-\frac{dk^0}{dk^i}x^0=0, \nonumber\\
\end{eqnarray} 
so, denoting by $v_g$, the group velocity of the wave packet, we get
\begin{equation}
v_g=\left|\frac{\delta x^i}{\delta x^0}\right|=\left|\frac{d k^0}{d k^i}\right|=1,
\end{equation} 
where in the last step we used relation (\ref{epot23}).

It is in point to stress that it is {\it not} the fluid which flows with the speed of light, rather it is the disturbances propagating in the flow field which travel with the above calculated group and phase velocities.

In previous work on the subject, there has been much debate as to the actual definition of incompressibility of relativistic fluids. The most straight forward definition in the framework of standard relativity, would be to demand that an element of the fluid shall retain the same proper volume throughout its motion, or in other words that the expansion of an elementary world-tube of liquid shall be zero. The difficulty lies in the fact that this definition yields infinite sound wave velocities. This would violate the most basic assumptions of standard relativity which forbids any material particle or information to be propagated at a speed higher than light. A disturbance such as a sound wave does carry energy, momentum and information and should therefore not break the light speed barrier. For this precise reason Synge \cite{Synge227} introduced two more alternate definitions, the second of which actually defines an incompressible fluid as such that sound waves propagate within it at light velocity.

In comparison, the equations we have written down and the simple assertion that incompressibility is to be defined on events in four dimensional space-time, lead smoothly and naturally to light speed sound waves in agreement with Synge's criterion. Moreover it can be seen that as $c\to\infty$, the sound waves speed, tends to infinity too, as would be expected of an incommpresible fluid in the nonrelativistic limit.

%%%%%%%%%%%%%%%%%%%%%%%%%%%%%%%%%%%%%%%%%%%%%%%%%%%%%%%%%
\section{Compressible fluid, Acoustic approximation}

We wish to widen the scope of our study to encompass the behavior of a compressible fluid, and we shall show that this additional freedom introduces acoustic modes with sound velocities other than that of light.

We begin by assuming that the density and pressure vary only slightly from some constant values. 

\begin{equation}\label{eacc1}
n=n_0+n',\qquad p=p_0+p'.
\end{equation}
Generally the pressure is a function of the density so one can expand the pressure in a Taylor series:

\begin{equation}\label{eacc2}
p(n)=p(n_0)+n'\frac{\partial p(n_0)}{\partial n}=p(n_0)+u^2n'
\end{equation}
where higher derivatives in $n$ have been neglected and we have in the last stage denoted the constant $u^2=\frac{dp(n_0)}{dn}$. Comparing equations (\ref{eacc1}) and (\ref{eacc2}) we get  a linear relationship between the variations of density and pressure,

\begin{equation}\label{eacc3}
p'=u^2n'.
\end{equation}
We wish also to assume the three space components of the velocity field to be, in some sense, small. The time component on the other hand must by definition be of the order of $c$ and can therefore not be considered small. This reasoning leads us to divide the velocity field too into two parts,
\begin{equation}\label{eacc4}
   v^\mu=v_0^\mu+v'^\mu,
\end{equation}
where $v_0^\mu=(v_0,0,0,0)$ is a constant\footnote{It is sufficient if $v_0$ varies only slowly in spacetime relative to $v'^\mu$; this possibility will be discussed later on.}, pure, time-like vector and $v'^\mu$ is a small perturbation the nature of which we wish to determine.
It has been shown earlier that the motion of the fluid, neglecting unimportant body forces, is governed by the following two equations: 		

\begin{eqnarray}
-\partial^\mu p=n\left( \frac{\partial v^\mu}{\partial\tau}+v^\nu\partial_\nu v^\mu\right)\label{eacc5}\\
-n\partial_\mu v^\mu=\frac{\partial n}{\partial\tau}+v^\mu\partial_\mu n\label{eacc6}
\end{eqnarray}
We assume firstly, that all changes in $\tau$ in equations (\ref{eacc5}) and (\ref{eacc6}), are negligably small compared with the spacetime changes, and we can therefore omit the $\tau$ derivatives from both equations. This can be explained too on grounds of integrating the above equations over $\tau$ and asserting that all physical quantities vanish as $\tau\to \pm \infty$ or by taking the zero component of the frequencies in $\tau$. All three arguments are essentially equivalent and are based on the fact that any apparatus in the laboratory is not capable of resolving fast changes in $\tau$, but can rather only measure averages over large periods of the universal time.  The resulting equations 
\begin{eqnarray}
-\partial^\mu p=nv^\nu\partial_\nu v^\mu,\label{eacc7}\\
-n\partial_\mu v^\mu=v^\mu\partial_\mu n,\label{eacc8}
\end{eqnarray}
can now be linearized  by inserting relations (\ref{eacc1}) and (\ref{eacc4}) and keeping only terms of the first order ($n'$, $p'$ and ${v'}^\mu$ being of the first order). Higher order terms are neglected and so in  this approximation the two equations reduce to

\begin{eqnarray}
-u^2\partial^\mu n'=n_0 v_0^\nu\partial_\nu v'^\mu,\label{eacc9}\\
-n_0\partial_\mu v'^\mu=v_0^\mu\partial_\mu n',\label{eacc10}
\end{eqnarray}
where by remembering relation (\ref{eacc3}), we have  eliminated  $p'$ from equation (\ref{eacc7}). The importance of the background vector  $v_0^\mu$, now becomes apparent. It ensures that the velocity field remains time-like and that the large time-like component does {\it not} get neglected in the linearization process.

By taking space derivatives of Eq. (\ref{eacc9}) it can be shown that	

\begin{equation}\label{eacc11}
v_0^\sigma\partial_\sigma(\partial^\mu v_\nu-\partial_\nu v^\mu)=0,
\end{equation}								
implying that if the flow was initially irrotational everywhere, then it will remain so throughout the entire evolution and that the velocity field can be derived from a potential:

\begin{equation}\label{eacc12}
v^\mu=-\partial^\mu \Phi.
\end{equation}
Multiplying Eq. (\ref{eacc9}) by $v_{0\mu}$ one obtains

\begin{equation}\label{eacc13}
-u^2 v_0^\mu\partial_\mu n'=n_0 v_{0\mu} v_0^\nu\partial_\nu v'^\mu,
\end{equation}
where we have interchanged the raising and lowering of the indices on the left hand side. The equations can now be solved for gradients of $v'^\mu$ by inserting Eq. (\ref{eacc10}) into (\ref{eacc13}):

\begin{equation}\label{eacc14}
u^2\partial_\mu v'^\mu=v_0^\nu v_{0\nu}\partial_\nu v'^\mu.
\end{equation}
Using relation (\ref{eacc12}) to replace the velocity field by the derivative of its potential $\Phi$, and rearranging terms, we get
 
\begin{equation}\label{eacc15}
\left(\partial_\mu-{v_0^\nu v_{0\mu}\over u^2}\partial_\nu\right)\partial^\mu\Phi=0,
\end{equation}
which can alternately be written as
 
\begin{equation}\label{eacc16a}
\left(g^\nu\,_\mu-U^\nu\,_\mu\right)\partial_\nu\partial^\mu\Phi=0,
\end{equation}
or
\begin{equation}\label{eacc16b}
T^\nu\,_\mu\partial_\nu\partial^\mu\Phi=0,
\end{equation}
where we denote the tensor $U^\nu\,_\mu={1\over u^2}v_0^\nu v_{0\mu}$ and $T^\nu\,_\mu=g^\nu\,_\mu-U^\nu\,_\mu$.

Before resuming with an interpretation of the result (\ref{eacc16a}), we wish to express the variation in density of events $n'$, in terms of the potential $\Phi$. Inserting relation (\ref{eacc12}) into Eq. (\ref{eacc9}) and rearranging terms we get

\begin{equation}\label{eacc17}
\partial^\mu\left(n'-{n_0\over u^2}v_0^\nu\partial_\nu\Phi\right)=0,
\end{equation}
implying that the term in parenthesis is constant. By definition though, any constant part of $n'$ is to be included in $n_0$, so the additive constant must be identically  zero and we are left with the following relationship
 
\begin{equation}\label{eacc18}
n'={n_0\over u^2}v_0^\nu\partial_\nu\Phi.
\end{equation}

We now return to Eq. (\ref{eacc16a}), and assume for simplicity that the background fluid is stationary i.e. $v_0^\mu=(v_0,0,0,0)$ is pure time-like and constant. We then get

\begin{equation}\label{eacc19}
\left[\partial_\mu\partial^\mu+\left({v_0^2\over u^2}\right)\partial_0\partial^0\right]\Phi=0, 
\end{equation}
or, on separating time and space derivatives,

\begin{equation}\label{eacc20}
\left[\nabla^2-\left(1+{v_0^2\over u^2}\right){1\over c^2}\frac{\partial^2}{\partial t^2}\right]\Phi=0.
\end{equation}
This is a wave equation for which the modified sound velocity is

\begin{equation}\label{eacc21}
\omega={cu \over \sqrt{u^2+ v^{02}}},
\end{equation}
and we can investigate the various limits as follows. When we take the fluid to its noncompressible limit, or mathematically $u\to\infty$, we get for the sound velocity $\omega\to c$. This agrees with our conclusions of the previous section, namely that a noncompressible fluid gives rise to light speed sound waves.

\begin{figure}[hbt]
\centerline{
   \epsfig{figure=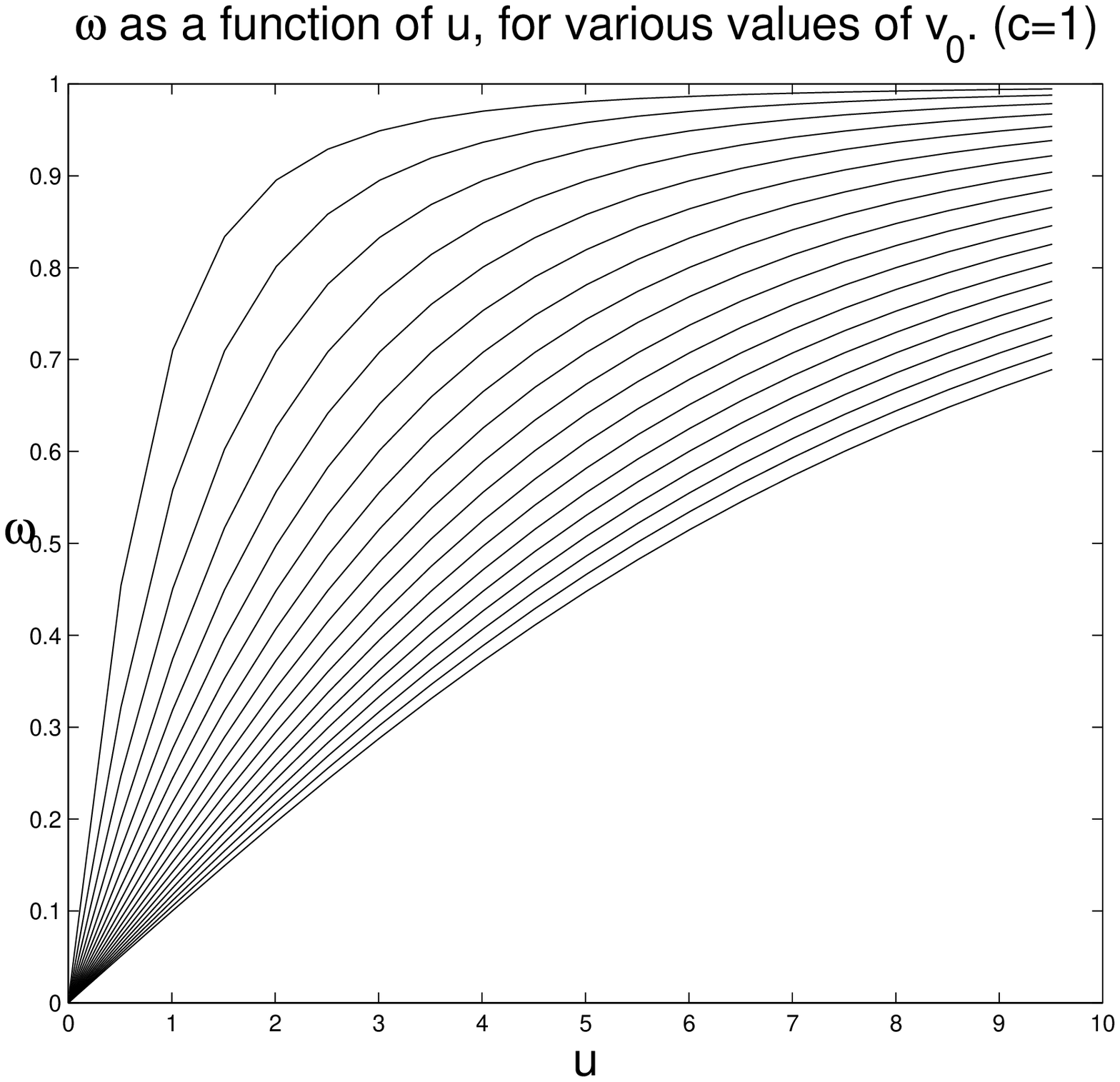, width=2.5in}
   \qquad
   \epsfig{figure=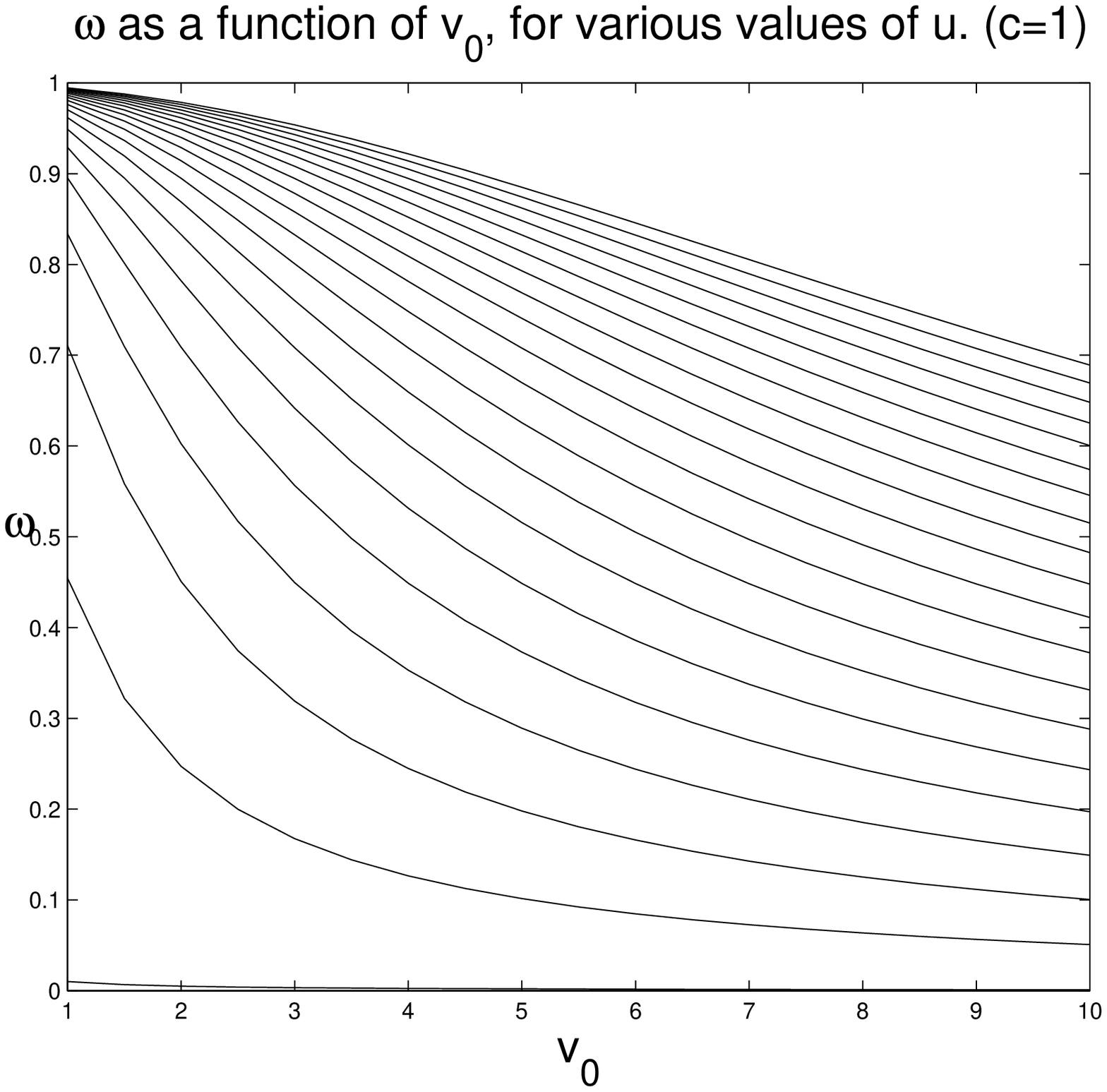, width=2.5in}
   }
   \caption[omega]{\label{omega}$\omega$ as a function of the inverse compressibility $u$, and energy $v_0$ respectively. }
\end{figure}

On the other hand if we take the fluid to be `sufficiently' compressible, or in other words $u \ll v_0$, the velocity becomes linear in $u$, $\omega\to {c\over v_0}u$. This is feasible for in the nonrelativistic limit when $v_0\to c$, we then get the usual nonrelativistic result for compressible flow namely, $\omega \to u$. These results are summarized in figure \ref{omega}.

%%%%%%%%%%%%%%%%%%%%%%%%%%%%%%%%%%%%%%%%%%%%%%%%%%%%%%%%%
\section{Viscous fluids}

Until now we have dealt solely with ``ideal fluids'' which do not sustain shear stresses during motion. The justification for our previous equations was that for many materials the shear stresses occurring during motion are small compared with the pressure. In practice though, all liquids and gases are in fact able to sustain shear forces, and we now wish to take these into account in a consistent covariant manner. 
In order to obtain equations describing the motion of a viscous fluid, we have to include some additional terms in the constitutive equations of the fluid. We therefore write the stress tensor $\sigma^{\mu\nu}$ in the form
\begin{equation}\label{evisc1}
 \sigma^{\mu\nu}=-pg^{\mu\nu}+\hat\sigma^{\mu\nu}.
\end{equation}
The second term is the extra stress tensor or viscosity stress tensor resulting from frictional forces between different layers of fluid. We can, with some restrictive assumptions, establish a general form for the tensor $\hat\sigma^{\mu\nu}$. If we assume that our fluid does not ``remember'' its past history or initial configuration but rather that its motion is governed solely by the immediately preceding state, then $\hat\sigma^{\mu\nu}$, must depend only on the velocity field of the fluid. Processes of internal friction occur in a fluid only when different fluid particles move with different velocities, so that there is a relative motion between various parts of the fluid. On the other hand if the distances between adjacent parts of the fluid are kept constant during the motion and assuming that forces between events are a function of the distance separating them $\Delta x_\mu \Delta x^\mu$, then there will be no friction.  In other words a frictional dissipation could arise only in those regions of the fluid continuum undergoing distortion, excluding places at which the fluid moves uniformly as a rigid body. Hence $\hat\sigma^{\mu\nu}$ must depend on derivatives of the velocity field rather than the velocity field itself.  When the velocity gradients are small we can, to some approximation, suppose  $\hat\sigma^{\mu\nu}$ to be a linear function of the derivatives $\partial^\mu v^\nu$ omitting higher orders of powers and derivatives. There can be no terms independent of $\partial^\mu v^\nu$, since $\hat\sigma^{\mu\nu}$ must vanish for constant $v^\mu$. 

A rigid rotation of the fluid in space time must also be excluded from affecting the viscous force, and we shall show that this implies that only the symmetric combination of derivatives $\partial^\mu v^\nu+ \partial^\nu v^\mu$, can be contained in  $\hat\sigma^{\mu\nu}$. We must note here that by rigid rotation in spacetime, we mean the group of transformations $\Lambda^{\mu\nu}$, which keep the covariant distance $dx_\mu dx^\mu$ constant. This is precisely the group of pure Lorentz transformations, which fulfill the requirement 

\begin{equation}\label{evisc1a}
\Lambda^\mu\,_\nu\Lambda_\gamma\,^\nu=g^\mu\,_\gamma.
\end{equation}
We can show quite generally that for an infinitesimal transformation $\Lambda^\mu\,_\nu=g^\mu\,_\nu+\varepsilon^\mu\,_\nu$, the infinitesimal displacement tensor $\varepsilon^\mu\,_\nu$, must be antisymmetric, as follows:

\begin{eqnarray}\label{evisc1b}
\Lambda^\mu\,_\nu\Lambda_\gamma\,^\nu&=&(g^\mu\,_\nu+\varepsilon^\mu\,_\nu)(g_\gamma\,^\nu+\varepsilon_\gamma\,^\nu)\nonumber\\
&=& g^\mu\,_\nu g_\gamma\,^\nu+\varepsilon^\mu\,_\nu g_\gamma\,^\nu+g^\mu\,_\nu\varepsilon_\gamma\,^\nu+O(\varepsilon^2)\nonumber\\
&\approx& g^\mu\,_\gamma+(\varepsilon^\mu\,_\gamma+\varepsilon_\gamma\,^\mu),
\end{eqnarray}
where in the third line we have neglected terms of second order in $\varepsilon$. Comparison of Eq. (\ref{evisc1b}) with (\ref{evisc1a}) yields

\begin{equation}\label{evisc1c}
\varepsilon^\mu\,_\gamma=-\varepsilon_\gamma\,^\mu,
\end{equation}
proving the antisymmetry of $\varepsilon^\mu\,_\gamma$.

\begin{figure}[h]
   \fbox{\epsfig{figure=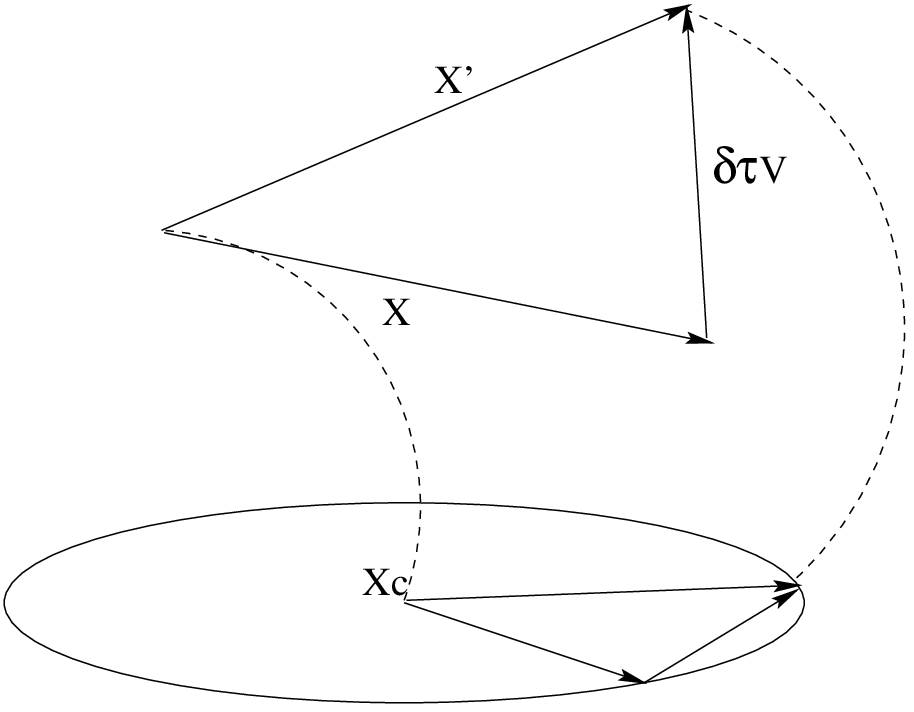, width=1.5in, angle=270}}
   \caption[rot]{\label{rot}Infinitesimal rotation of fluid.}
\end{figure}

Consider now a small region in the flow around a point $x_c^\mu$, at which the fluid can locally be considered to be rotating rigidly (figure \ref{rot}). Our claim is that at $x_c^\mu$ there are no frictional forces within the fluid and therefore the the viscosity stress tensor must vanish at that point. Let $\Omega^\mu\,_\nu$ represent the angular velocity of the fluid rotating in the $\mu-\nu$ plane. The velocity field round $x_c^\mu$ will then by definition be

\begin{equation}\label{evisc2}
v^\mu=\Omega^\mu\,_\nu x^\nu,
\end{equation}
where $x^\nu$ here and in the following discussion is the displacement from the center of rotation $x_c^\mu$. Now consider a fluid particle at some place $x^\mu$ being rotated  by an infinitesimal amount to a new position $x'^\mu$. There are two relations connecting $x$ and $x'$:

\begin{eqnarray}\label{evisc3}
x'^\mu&=&\Lambda^\mu\,_\nu x^\nu\label{evisc4},\\
x'^\nu&=&x^\nu+\delta \tau v^\nu\label{evisc5},
\end{eqnarray}
where 

\begin{equation}\label{evisc6}
\Lambda^\mu\,_\nu=g^\mu\,_\nu+\varepsilon^\mu\,_\nu,
\end{equation} 
is an infinitesimal Lorenz transformation. 
Inserting Eq. (\ref{evisc2}) into (\ref{evisc5}), and comparing with Eq. (\ref{evisc4}) results in

\begin{equation}\label{evisc7}
\Lambda^\mu\,_\nu=g^\mu\,_\nu+\delta \tau\Omega^\mu\,_\nu 
\end{equation}
Using relation (\ref{evisc6}) we can get a form for $\Omega^\mu\,_\nu$ in terms of $\varepsilon^\mu\,_\nu$

\begin{equation}\label{evisc8}
\delta \tau\Omega^\mu\,_\nu=\varepsilon^\mu\,_\nu,
\end{equation}
proving that $\Omega^\mu\,_\nu$ too, is antisymmetric. Taking derivatives of relation (\ref{evisc2}) while considering the fact that $\Omega^\mu\,_\nu$ is antisymmetric shows that the combination $\partial^\mu v^\nu+ \partial^\nu v^\mu$ vanishes for a rigidly rotating fluid whereas the antisymmetric combination $\partial^\mu v^\nu- \partial^\nu v^\mu$, does not. As  the viscosity stress tensor  $\hat\sigma^{\mu\nu}$ must vanish for rigidly rotating motion, we deduce that it must contain just the symmetric combination of derivatives.

\begin{figure}[h]
\centerline{
   \fbox{\epsfig{figure=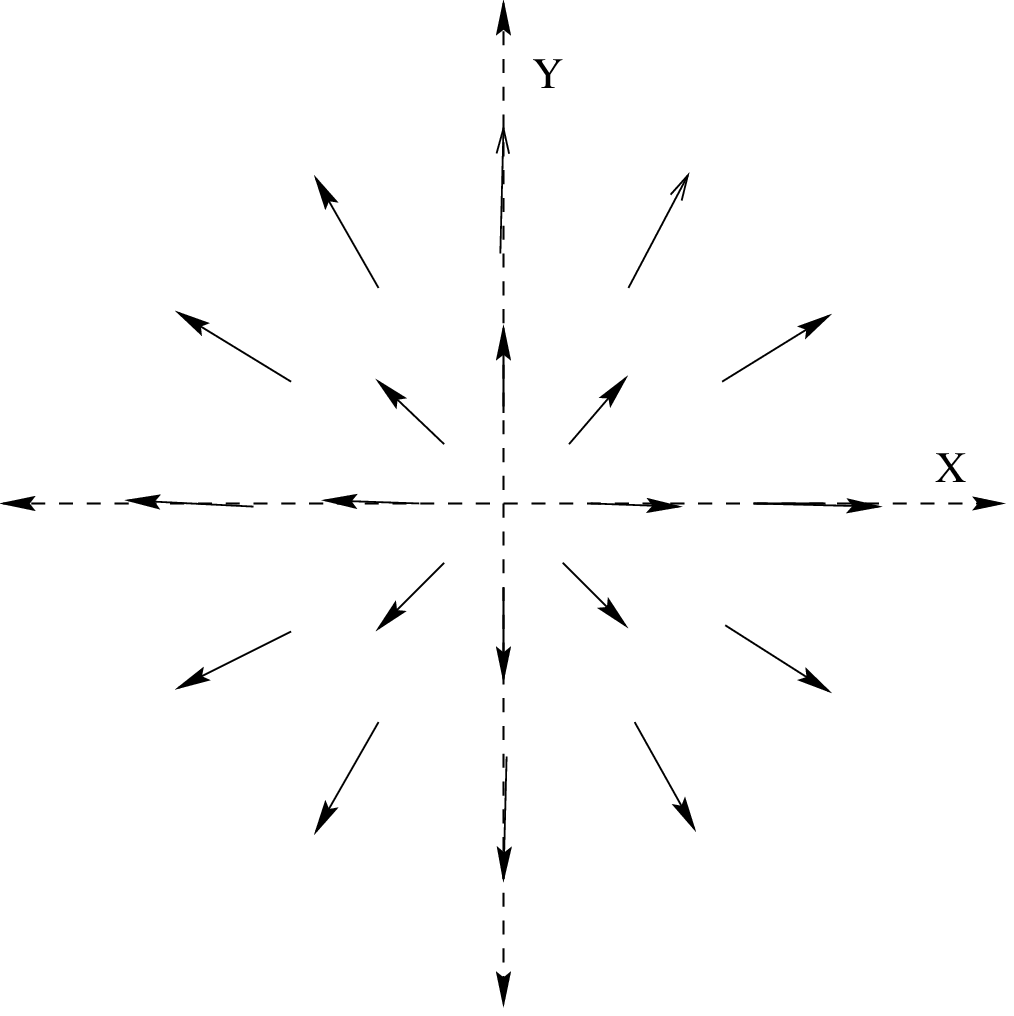, width=1.5in, angle=270}}
   \qquad
   \fbox{\epsfig{figure=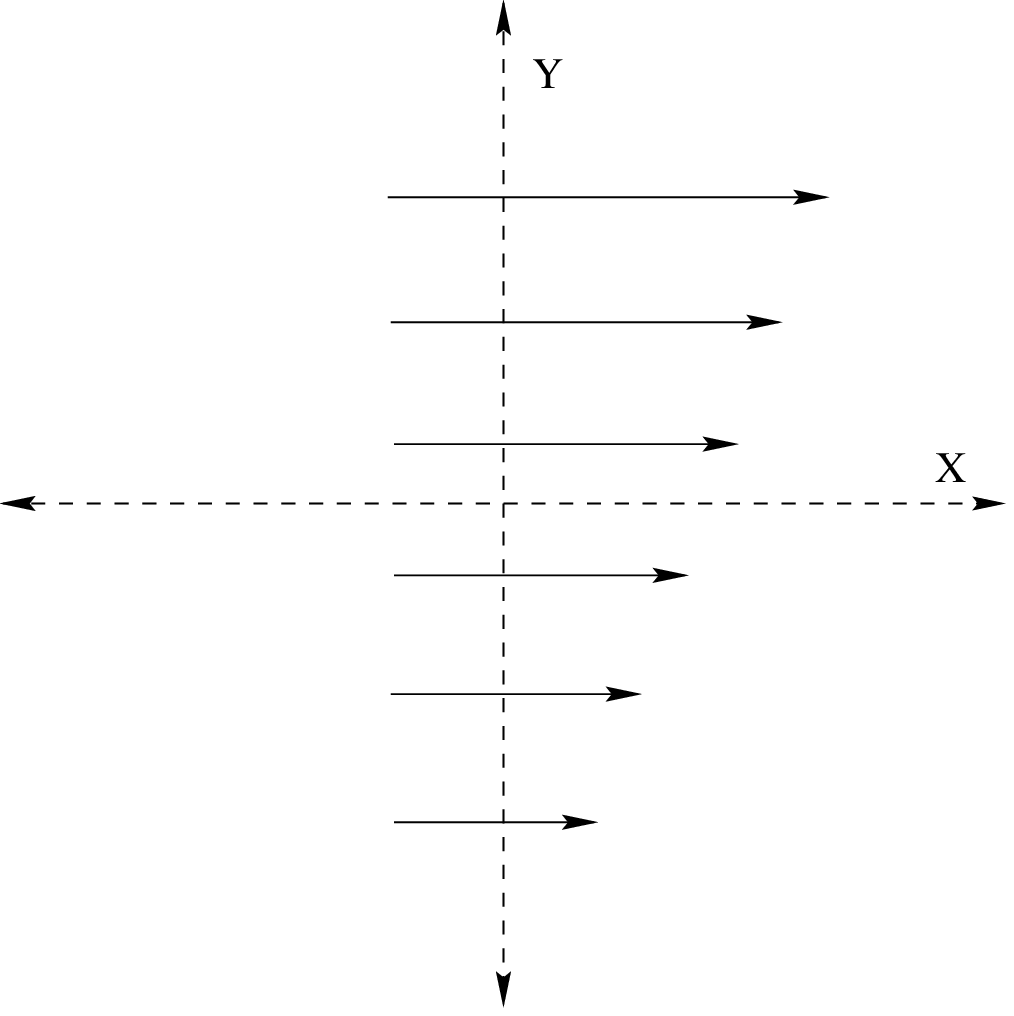, width=1.5in, angle=270}}
   }
   \caption[distort]{\label{distort} Two basic distortions: {\it left:} Radial flow. {\it right} Shear flow.}
\end{figure}

The most general tensor of rank two satisfying all the above conditions is
\begin{equation}\label{evisc9}
\hat\sigma^{\mu\nu}=a(\partial^\mu v^\nu+\partial^\nu v^\mu)+b g^{\mu\nu}\partial_\gamma v^\gamma,
\end{equation}
where $a$ and $b$ are independent of the velocity. It is convenient, however, to replace $a$ and $b$ by other constants and write the equation in another form which lends itself more readily to interpretation

\begin{equation}\label{evisc10}
\hat\sigma^{\mu\nu}=\eta(\partial^\mu v^\nu+\partial^\nu v^\mu-\frac{1}{2} g^{\mu\nu}\partial_\gamma v^\gamma)+\zeta g^{\mu\nu}\partial_\gamma v^\gamma.
\end{equation} 
We shall call the constants $\eta$ and $\zeta$ the shear viscosity and bulk viscosity respectively. The terminology derives from considering the two following basic flows (see figure \ref{distort}):

\begin{enumerate}

\item 
For a ``pure'' shear flow $v^\mu=u(n'_\gamma x^\gamma)n^\mu$, where $n^\mu$ and $n'^\mu$ are mutually orthogonal unit vectors so that the direction of the velocity gradient $n'^\mu$ is perpendicular to the direction $n^\mu$ of the velocity, it can be shown that only the first term in $\hat\sigma^{\mu\nu}$ survives, leaving  $\hat\sigma^{\mu\nu}=u\eta(n^\mu n'^\nu+n'^\mu n^\nu)$. So the coefficient $\eta$ alone governs the resistance of the fluid to shear distortions.  

\item 
For a ``pure'' radial flow $v^\mu=(u/4)x^\mu$ in which the material expands in a radial direction, the the first term in Eq. (\ref{evisc10}) vanishes leaving  $\hat\sigma^{\mu\nu}=u\zeta g^{\mu\nu}$, which shows that $\zeta$ is proportional to the resistance of the fluid to expansion of bulk. This can be understood also by noticing that the expression in parentheses in Eq. (\ref{evisc10}) has the property of vanishing on contraction with respect to $\mu$ and $\nu$.

\end{enumerate}

The equations of motion can now be obtained by adding the expressions $\partial_\nu\hat\sigma^{\mu\nu}$ to the left hand side of the nonviscous flow equation (\ref{eacc5}). Thus we have 

\begin{equation}\label{evisc11}
-\partial^\mu p+\partial_\nu\left\{\eta\left(\partial^\mu v^\nu+\partial^\nu v^\mu-\frac{1}{2} g^{\mu\nu}\partial_\gamma v^\gamma\right)+\zeta g^{\mu\nu}\partial_\gamma v^\gamma\right\}=n\left(\frac{\partial v^\mu}{\partial \tau}+v^\nu\partial_\nu v^\mu\right).
\end{equation}
This is the most general form of the equations of motion of a viscous fluid. If we assume, however, that the viscosity coefficients do not change noticeably throughout the fluid, then they may be regarded as constant and can therefore be taken outside the gradient operators. We then have

\begin{equation}\label{evisc12}
-\partial^\mu p+\eta \partial_\nu\partial^\nu v^\mu+\left(\zeta+\frac{1}{2}\eta\right)\partial^\mu\partial_\nu v^\nu
=n\left(\frac{\partial v^\mu}{\partial \tau}+v^\nu\partial_\nu v^\mu\right).
\end{equation}    
Further simplification can be obtained by assuming the bulk viscosity $\zeta$ small compared to the shear viscosity $\eta$.

If the fluid is incompressible then $\partial_\nu v^\nu=0$ and the third term on the left  hand side of Eq. (\ref{evisc12}) vanishes giving

\begin{equation}\label{evisc13}
-\partial^\mu p+\eta \partial_\nu\partial^\nu v^\mu
=n\left(\frac{\partial v^\mu}{\partial \tau}+v^\nu\partial_\nu v^\mu\right).
\end{equation}  

On solving a specific dynamical problem one must also write down the boundary conditions for the equations of motion of a viscous fluid. We assume that there are friction forces acting between the boundary surface and the fluid such that the layer of fluid immediately adjacent to the boundary is brought to complete rest. Accordingly, the boundary conditions on the equations of motion require that the fluid velocity should vanish at fixed solid surfaces:
\begin{equation}\label{evisc14}
v^\mu=0.
\end{equation}
We emphasize that for a viscous fluid both the normal and tangential velocity components must vanish, in contradistinction to ideal fluids for which it is required only that the normal component vanish. In a general case with boundaries moving in $\tau$, the velocity of the fluid at the boundary should be equal to the velocity of the moving surface.

%%%%%%%%%%%%%%%%%%%%%%%%%%%%%%%%%%%%%%%%%%%%%%%%%%%%%%%%%
\section{Conclusions}

We have studied a continuum flow of events in space-time parameterized by $\tau$, the invariant, universal, historical time. In the process of evolution, as $\tau$ changes uniformly, the events move in space-time, generating a dense continuum of world lines which constitute a physical flow of particles. The velocity field is in general a function of both space-time and $\tau$ and is governed by equations of motion and continuity. The solution of the equations for the velocity field provides, in principal, a solution to the physical problem, but is in general rather complicated.
 
We considered an inviscid fluid and derived an analogue of the Bernoulli equation. In the special case of irrotational flow we proved that the velocity field can be derived by a potential. If in addition, the fluid is incompressible, it has been shown that the potential must obey the d'Alembert equation, and thus the problem is reduced to solving the d'Alembert equation with specific boundary conditions. The solutions consist of a background flow (The nonrelativistic solution) over which wave like ripples propagate with group and phase velocities equal to that of light. This is to be compared with Synge \cite{Synge276}, who actually defines noncompressibility in such a way as to achieve this result.

Relaxing the constraints and allowing the fluid to become linearly compressible one can derive a wave equation, from which the sound velocity can again be computed. For a stationary background flow, it has been demonstrated that the sound velocity, attains its correct values for the incompressible and nonrelativistic limits. It is in principal possible, under some restrictions, to compute the sound velocities also for a non stationary and non uniform background flow. Precise computations and predictions in this direction are yet to be studied.   

Finally viscosity was introduced and the use of some general arguments of symmetry and isotropy enabled the formulation of equations for the motion of a viscous fluid. Only introductory comments were made about this topic and the equations derived are yet to be applied to specific problems. 
 
It must be emphasized that the velocity and density (and any other) fields that we mention, describe quantities related to the flow of events at a particular time $\tau$. In order to derive a physically observable quantity related to particles, an averaging must be performed over $\tau$. The method we use is that of integrating over $\tau$ with the density of events as a weight function. An exact transcription from event properties to world-line, or particle, properties, however, is not always possible due to nonlinearities of these quantities. Alternatively the observed quantities can be extracted from the zero frequency (in $\tau$) components of the fields.
%%%%%%%%%%%%%%%%%%%%%%%%%%%%%%%%%%%%%%%%%%%%%%%%%%%%%%%%%
\bibliographystyle{plain}
%\bibliographystyle{aip}

%%%%%%%%%%%%%%%%%%%%%%%%%%%%%%%%%%%%%%%%%%%%%%%%%%%%%%%%%%%%%%%

%\renewcommand{\baselinestretch}{1}

\newpage

\newpage 

\newpage

\newpage

{$\omega$ as a function of the inverse compressibility $u$, and energy $v_0$ respectively.}

\vspace{1in}
   {Infinitesimal rotation of fluid.}

\vspace{1in}
   {Two basic distortions: {\it left:} Radial flow. {\it right} Shear flow.}
\end{document}